\documentstyle[epsfig,11pt]{article}

\def\be{\begin{equation}} \def\ee{\end{equation}} \def\bea{\begin{eqnarray}}
\def\eea{\end{eqnarray}} \def\nnb{\nonumber}

\begin{document}

\hfill{\tt January 10, 2001}

\hfill{\tt SNUTP 99-052, TRI-PP-00-01, USC(NT)-Report-00-1}

\begin{center}
{\Large {\bf Threshold $pp \to pp\pi^0$ up to one-loop accuracy}}

\vskip 5mm

{\large
Shung-ichi Ando$^{a,}$\footnote{E-mail:sando@nuc003.psc.sc.edu},
Tae-Sun Park$^{b,}$\footnote{E-mail:tspark@nuc003.psc.sc.edu},
and Dong-Pil Min$^{c,}$\footnote{E-mail:dpmin@phya.snu.ac.kr}}

\vskip 5mm

{\large {\it $^{a}$Department of Physics and Astronomy,
University of South Carolina, \\
Columbia, South Carolina 29208}}

\vskip 1mm

{\large {\it $^{b}$Theory group, TRIUMF, Vancouver, B. C., Canada V6T 2A3}}

\vskip 1mm

{\large {\it $^{c}$Department of Physics,
Seoul National University, Seoul 151-742, Korea}}

\end{center}

\vskip 1cm

The $pp\to pp\pi^0$ cross section near threshold is computed up to
one-loop order including the initial and final state interactions
using the {\it hybrid} heavy baryon chiral perturbation theory and
the counting rule {\it a la} Weinberg. With the counter terms
whose coefficients are fixed by the resonance-saturation
assumption, we find that the one-loop contributions are 
as important as the tree-order contribution 
and bring the present theoretical
estimation of the total cross section close to the experimental
data. The short-ranged contributions are controlled by means of a
cutoff, and a mild cutoff dependence is observed when all
diagrams of the given chiral order are summed. 
To the order treated, however, 
the expansion is found to converge rather
slowly, calling for further studies of the process.


\renewcommand{\thefootnote}{\#\arabic{footnote}}
\setcounter{footnote}{0}

\newpage

\noindent {\bf 1. Introduction}

The accurate measurements \cite{exp1,exp2}
of the total cross section near threshold of the process
\be
p+p\rightarrow p+p+\pi^0 
\label{pppi0} 
\ee 
have stimulated many
theoretical investigations \cite{HME, off, tree;bypark,
tree;cohen, tree;sato, tree;kolck, loop;gedalin, loop;DKMS,
gedalin2,bernard}, but a theoretical explanation of the data is
not yet completed. The difficulties in describing the process from
the low-energy effective field theories such as the heavy baryon
chiral perturbation theory (HBChPT) \cite{hbchpt} are twofold:
Firstly, the leading tree order contributions of both the impulse
(IA) and the meson-exchange (MEC) diagrams are suppressed and the
contributions from sub-leading IA and MEC diagrams are almost
canceled off. Secondly, the momentum of the process at threshold
is of the scale of $\sim\sqrt{m_\pi m_N}$ where $m_\pi$ and $m_N$
are the pion and proton mass, respectively, which is considerably
bigger than the usual characteristic scale, $m_\pi$. Thus
contributions from higher order operators -- which in general are
controlled only poorly -- can be of non-negligible importance
calling for the calculation of higher order contributions in the
expansion. These are the two principal reasons for questioning 
the predictive power of the HBChPT approach 
(in its standard form of the Weinberg's counting rule\cite{weinberg})
to the reaction in question.

According to the chiral filter
mechanism\cite{chiralfilter,chiralfilter2}, 
the processes such
as isovector M1 transitions and axial-charge weak transitions,
which
are dominated by one-soft-pion exchange terms 
(i.e., current algebra), 
are chiral-filter-protected so that corrections to the
leading order terms are suppressed and can be systematically
controlled by chiral perturbation theory. 
When a process is {\it not} dominated 
by soft-pions for reasons of symmetry and/or
kinematics, then it is unprotected by the chiral filter and
consequently higher order terms (involving short-range ones)
become non-negligible, making a systematic chiral expansion
difficult, if not impossible. 
The suppression of the leading order contribution and the 
substantially cancellation between the sub-leading contributions
of the process (\ref{pppi0}) make
this process quite similar to other chiral-filter unprotected
cases such as the isoscalar M1 and E2 matrix elements in the
polarized $np$-capture process, $\vec{n}+\vec{p}\rightarrow
d+\gamma$, 
discussed in \cite{chiralfilter2} and the solar ``hep"
process studied in \cite{PKMRhep}. 
However unlike these processes
which can be calculated with some confidence because of
the small momentum involved, the $\pi^0$ production process which
involves a relatively large momentum has the second (kinematic)
condition which makes the calculation even more difficult. 
We hope
in this paper to shed some lights on these issues which have not
been fully  explored up to date.

HBChPT is a consistent and systematic low-energy effective field
theory, whose expansion parameter is $Q/\Lambda_\chi$, where $Q$
is the typical momentum scale involved and/or pion mass, while
$\Lambda_\chi\sim m_N\sim 4\pi f_\pi$ is the chiral scale,
$f_\pi\simeq 93$ MeV the pion decay constant. Due to the large
momentum scale of the process, $Q\sim \sqrt{m_\pi m_N}$, the
convergence of the chiral expansion \cite{tree;cohen, tree;kolck}
and non-relativistic treatment \cite{gedalin2,bernard} of the
process have been questioned. 
However, there have been many
attempts to apply the theory because the scale $Q$ is still
smaller than $\Lambda_\chi$, ({\it e.g.},
$Q/\Lambda_\chi\sim\sqrt{m_\pi/m_N}\simeq 0.4)$. Some tree-order
calculations \cite{tree;bypark,tree;cohen,tree;sato,tree;kolck}
and partial one-loop calculations \cite{loop;gedalin} have been
reported. Quite recently, Dmitra\v{s}inovi\'{c} {\it et al.}
\cite{loop;DKMS} analyzed all relevant transition operators of
one-loop order and found that important contributions may rise
from the one-loop diagrams. In this paper, we report the result of
our calculation of the reaction cross section by using ``hybrid
approach"\cite{PMR} with those transition operators including the
initial and final state interactions (ISI and FSI) , and
short-ranged contributions whose coefficients are determined
through the resonance-saturation assumption \cite{resonance}.
Indeed, we confirm here that the one-loop contribution is actually
quite important, which is not surprising for those processes that
are chiral-filter-unprotected~\cite{chiralfilter2,PKMRhep}.
Whether or not a reliable prediction can be made in the EFT
framework, even though the loop-order corrections are not
negligible -- as in the case of the {\it np}-capture and the
``hep" process -- will be the subject of our discussion.\\

\noindent {\bf 2. Hybrid approach for the pion production}

\begin{figure}[th] %
\begin{center}
\epsfig{file=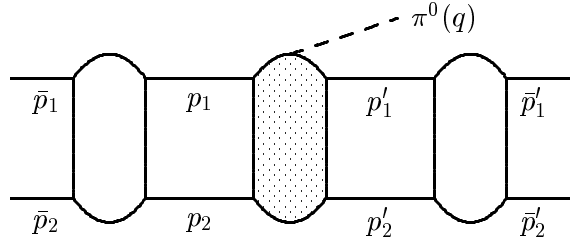}
\caption{\it Generic diagram of the
$pp\rightarrow pp\pi^0$. The empty blobs represent initial- and
final-state interactions (ISI and FSI), 
and the shaded blob the two-nucleon
irreducible transition operators.
The momenta in the asymptotic region
have bars attached to them.} \label{generic}
\end{center}
\end{figure}
The generic diagram for the process is drawn in 
Fig. \ref{generic}. 
In calculating the process, we adopt the so-called ``hybrid
approach'', where the two-nucleon irreducible transition operators
(shaded blob in the figure) are obtained within HBChPT, and all
the reducible parts are embedded into the phenomenological
wavefunctions. The justification for this procedure as a viable
EFT was given \cite{chiralfilter2, PKMRhep, HPM} and will not be
expounded here again. We note that it is consistent with the
spirit of Weinberg's original proposal of applying EFT to nuclear
physics\cite{weinberg}. 
It is well known that, near threshold, the
reaction is dominated by the transition between $|{}^3P_0\rangle$
(initial) and $|{}^1S_0\rangle$ (final) states. 
With other partial contributions neglected, 
the shape of the total cross section is
well reproduced by taking into account only the phase factor and
the final state interaction \cite{exp1}. 
So we calculate the transition amplitude at the threshold kinematics,
$q^\mu=(m_\pi,\vec{0})$ for the four momentum of the emitted
$\pi^0$. According to Sato {\it et al.} \cite{tree;sato}, this
approximation reduces the cross sections by up to 10 \% in the
range $0\le |\vec{q}|\le 0.4\ m_\pi$.

For the discussion on the kinematics, we denote the incoming
(outgoing) momentum of the $j$-th proton by $p_j^\mu$
(${p'}_j^\mu$).
In terms of the relative momentum $\vec{p}$ and $\vec{p'}$,
they can be written as
$\vec{p}_1= \vec{p}$,
$\vec{p}_2= -\vec{p}$,
$\vec{p'}_1= \vec{p'}$ and
$\vec{p'}_2= -\vec{p'}$ in the CM frame.
For convenience, we also define the momenta transferred,
$k_j^\mu\equiv (p_j-p_j')^\mu$.
The momentum conservation then reads
\be
k_1^\mu+k_2^\mu=q^\mu=(m_\pi,\,\vec{0}). \label{eq;con} \ee
Hereafter we will put bars on the momenta defined in the
asymptotic region where particles are on-shell\footnote{
The on-shell condition in HBChPT is given by
$p_{j,{\rm rel}}^2-m_N^2 = p_j^2 + 2 m_N v\cdot p_j=0$, 
where 
$p_{j,{\rm rel}}^\mu$ is the usual (relativistic) four-momentum:
$p_{j,{\rm rel}}^\mu \equiv m_N v^\mu + p_j^\mu$ where 
$v^\mu$ is a constant vector with $v^2=1$.
};
$|\vec{\bar k}_j|=|\vec{\bar p}|=\sqrt{m_\pi m_N + m_\pi^2/4}$,
$|\vec{\bar p'}|=0$ and ${\bar k}_j^0=m_\pi/2$ for $j=(1,\,2)$.

The transition operator depends on the energy-transfer,
$k_j^0\equiv p_j^0 - {p'_j}^0$, 
which brings up 
an ``off-shell'' ambiguity \cite{tree;bypark,tree;sato}.
A common method (used in this work as well) is to assume the
``fixed kinematics approximation'' (FKA), which replaces $k_j^0$
by the on-shell energy transfer ${\bar k}_j^0=\frac{m_\pi}{2}$. On
the other hand, it has been pointed out that different treatments
of the off-shell behavior can produce significant differences. 
Sato {\it et al.} \cite{tree;sato} suggested to use, 
instead of FKA, the equation of motion approximation (EMA), 
replacing $p^0$ by ${\vec{p}}^2/2 m_N$ and similarly for ${p'}^0$; 
thus $k_{1,{\rm EMA}}^0 = k_{2,{\rm EMA}}^0 = 
\frac{{\vec{p}}^2 - {\vec{p'}}^2}{2 m_N}$. 
With this assumption, they carried out a momentum-space
calculation and obtained quite different results from those
obtained with FKA. While the EMA can be regarded as an approach
containing more dynamics than the FKA, neither is free from
drawbacks. 
For instance, it is to be noted that the EMA violates the
energy conservation, Eq. (\ref{eq;con}).

\noindent {\bf 3. The total cross section up to one-loop order}

Weinberg's counting rule \cite{weinberg} dictates that a Feynman
diagram 
with $\nu=2(1-C) + 2 L + \sum_i \bar \nu_i$
is of order of $(Q/\Lambda_\chi)^\nu$
where $C$ ($L$) is the number of the separate pieces (loops)
and the index $i$ runs for all vertices.
The ${\bar \nu}_i$ denotes the chiral order of the $i$-th vertex,
${\bar \nu}\equiv d+ n/2 -2$,
where $n$ is the number of nucleon lines and
$d$ the number of derivatives or powers of $m_\pi$
involved in a vertex.
As the momentum scale encountered in the present reaction is
$\sqrt{m_\pi m_N}$, a separate counting on ($m_\pi/\Lambda_\chi$)
and ($\sqrt{m_\pi m_N}/\Lambda_\chi$) can be envisaged as the
order counting for better convergence, as proposed by Cohen {\it
et al.} \cite{tree;cohen,hanhart}.
We continue to use Weinberg's counting rule 
because $m_\pi/\Lambda_\chi$ is still smaller than one.

Here we wish to remark about the nucleon propagator.
Since the nucleon kinetic term, ${\vec p}^2/(2 m_N)$,
is of order of $m_\pi$,
one may think that it should be included in the leading-order
nucleon propagator, $1/(p^0 + i \epsilon)$.
However, this is not the case
for {\em irreducible} loops\footnote{
For reducible diagrams,
we have $p^0 \sim \vec{p}^2/(2m_N)$.
This means that we should include the kinetic
term in treating the reducible loops,
which we always do in solving Schr\"odinger equations.
In other words, the fact that $p^0 \sim \vec{p}^2/(2m_N)$ in reducible
diagrams is a general property of low-energy
two- and many-body systems
(one can recall the virial theorem here),
not a particular consequence of the large
momentum scale.}.
The energy component of the loop momentum $p^0$
can pick up a pole of the pion propagator, which has the
large momentum scale, and leads to $p^0\sim \sqrt{m_\pi m_N}$.
As a result, $p^0 \gg {\vec p}^2/(2 m_N)$,
and therefore the kinetic energy term can be treated perturbatively.

\begin{figure}[ht]
\begin{center}
\epsfig{file=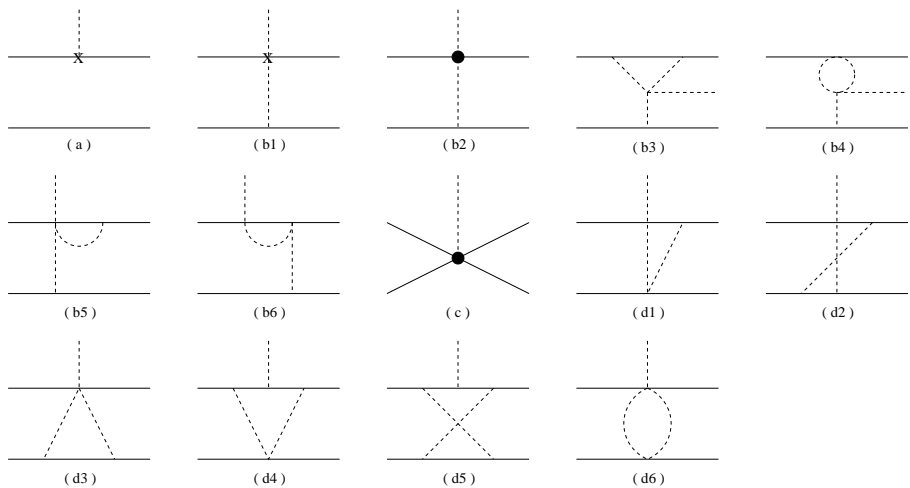,width=12cm }
\caption{\it Diagrams for the
pion production up to one-loop order.
The cross diagrams of (b5), (b6), (d1) and (d2) are omitted.
}
\label{fig;loops}
\end{center}
\end{figure}

Feynman graphs relevant to the process are drawn in Fig.
\ref{fig;loops}, where vertices of $\bar{\nu}=1$ are marked by
``X'', and those of $\bar{\nu}=2$ by a filled circle.
Due to the fact that, 
since the transition operator in between 
the initial state interaction (ISI) and the final state interaction (FSI)
is of off-shell,
a loop integral can have an off-shell singularity 
that breaks the symmetry \cite{tataru}.
We therefore employ the 
background field method (BFM) \cite{ecker}
which preserves symmetries
even for off-shell quantities such as the Green's function
and effective potential.
Among the 14 diagrams, eight graphs (a, b1, b2, b3, b6, c, d1, d2) 
contribute here: Graphs (b4), (d3), (d4), (d5) vanish at threshold,
(b5) is purely isovector-vector in the sea-gull vertex,
and the contribution from (d6) is identically zero due to isospin
symmetry. 
Note that (b3), (b4), (b5), (d3), (d6) depends on
the representation of pion field,
but their sum does not \cite{bernard-NpitoNpipi}.
We have not considered the recoil diagrams
(Fig. 4 in Ref.\cite{tree;cohen}), which are reducible
in our scheme and to be absorbed in the wavefunctions.
It has been reported that the recoil contribution
is small \cite{tree;cohen}.\footnote{\protect
In time-ordered perturbation theory (TOPT) the recoil diagrams
appear as irreducible.
The leading order recoil contribution should be, however,
removed provided that we are using the usual static one-pion-exchange 
(OPE) potential instead of the non-local one of the TOPT,
as Weinberg observed in connection to the 
three-nucleon potential \cite{weinberg}.
The recoil contribution given in Ref. \cite{tree;cohen} is one higher order
than the mentioned leading order one, and can be obtained by the
reducible OPE diagrams (Fig. $2(a,b)$ of Ref. \cite{tree;cohen})
with Feynman pion propagators, subtracted by the same diagrams
with static pion propagators.
Imposing the asymptotic condition for the external lines
of the diagram, however, one can show that
the recoil diagrams give us vanishing contribution at this order.
It then implies that the recoil contribution given in \cite{tree;cohen}
is reducible in TOPT as well as in our covariant scheme.
Thus including the contribution causes a double counting
in both schemes, provided that all the reducible diagrams
are embodied in the wavefunctions.
}

The HBChPT Lagrangian is expanded as
\bea {\cal L} = \sum{\cal L}_{\bar{\nu}}
={\cal L}_0 + {\cal L}_1 + {\cal L}_2+\cdots .
\eea
The Lagrangian relevant to our calculation reads
\bea
{\cal L}_0 &=& \bar{N}[iv\cdot D+2ig_AS\cdot \Delta]N
+ f_\pi^2{\rm Tr}\left(i\Delta^\mu i\Delta_\mu+
\frac{\chi_+}{4}\right),
\\
{\cal L}_1 &=& \bar{N}\left[ \frac{-D^2}{2m_N}
+\frac{g_A}{m_N}\{v\cdot \Delta,S\cdot D\} +c_1{\rm
Tr}(\chi_+)+\left(\frac{g_A^2}{2m_N}-4c_2\right)(v\cdot \Delta)^2
-4c_3\Delta\cdot \Delta\right] N , \nnb \\ \label{eq;nlo}
\\
{\cal L}_2 &=&
\bar{N}\frac{i}{2m_N}\left(\frac{g_A^2}{2m_N}v\cdot \Delta\Delta\cdot D
-4c_2{\rm Tr}(v\cdot\Delta\Delta^\mu)D_\mu\right)N+{\rm h.c.}
\nnb \\ &&
+\frac{g_A}{(4\pi f_\pi)^2f_\pi^2}
\left\{d_1^{(2)}\bar{N}[v\cdot\Delta S\cdot D-S\cdot\stackrel{\leftarrow}{D}
v\cdot\Delta]N\bar{N}N
\right.  \nnb \\ &&
+d^{(2)}_2\bar{N}v\cdot \Delta S^\mu N\bar{N}
[D_\mu-\stackrel{\leftarrow}{D}_\mu]N
+d^{(2)}_3\bar{N}
[v\cdot \Delta D_\mu-\stackrel{\leftarrow}{D}_\mu v\cdot \Delta]N\bar{N}S^\mu N
\nnb \\ &&
+d^{(2)}_4\bar{N}v\cdot \Delta N\bar{N}
[S\cdot D-S\cdot\stackrel{\leftarrow}{D}] N
+d^{(2)}_9 i\epsilon^{\alpha\beta\mu\nu}v_\alpha[\bar{N}v\cdot\Delta S_\mu
D_\beta N\bar{N}S_\nu N
\nnb \\ &&
\left.
+\bar{N}\stackrel{\leftarrow}{D}_\beta v\cdot\Delta S_\mu N\bar{N}S_\nu N
-\bar{N}v\cdot \Delta S_\mu N\bar{N}S_\nu D_\beta N
-\bar{N}v\cdot \Delta S_\mu N\bar{N}\stackrel{\leftarrow}{D}_\beta S_\nu N]
\right\} ,
\label{eq;nnlo}
\eea
where, in the absence of external fields,
$D_\mu = \partial_\mu+\Gamma_\mu$,
$\Gamma_\mu = \frac{1}{2}[\xi^\dagger,\partial_\mu \xi]$,
$\Delta_\mu = \frac{1}{2}\{\xi^\dagger,\partial_\mu \xi\}$,
$\chi_+=\xi^\dagger\chi\xi^\dagger+\xi\chi^\dagger\xi$,
$\chi = m_\pi^2$ and
$\xi = {\rm exp}\left(i\frac{\vec{\tau}\cdot \vec{\pi}}{2f_\pi}\right)$;
$v^\mu=(1,\vec{0})$ is the four-velocity vector,
$S^\mu=(0,\frac{\vec{\sigma}}{2})$ is the spin operator
and $g_A$ is the axial-vector coupling constant.
$c_i$ and $d^{(2)}_i$ are low energy constants which cannot be fixed
by the symmetry.
The values of $d^{(2)}_i$ are fixed by using resonance saturation below.
For the values of the $c$'s we use those obtained by Bernard
{\it et al.} \cite{bernard-loop}
\footnote{
%
The values of $c_i$'s have been updated
in \cite{tree;bypark,tree;cohen}.
We have checked that our results do not depend on these parameter set
much, less than $\simeq 10\ \%$ in total cross section.
},
\bea
c_1=-0.93\pm 0.10,\ \  c_2=3.34\pm
0.20, \ \ c_3=-5.29\pm 0.25\ [{\rm GeV}^{-1}] .
\label{eq;LECs-loop}
\eea
Note that all the degrees of freedom other than nucleons and pions
have been integrated out from the Lagrangian, their
roles encoded in the coefficients of the Lagrangian. 
In fact, there has been a claim that 
the direct treatment of $\Delta(1232)$ is important
\cite{tree;cohen}, and the two-body $\Delta$-contribution 
substantially cancels the leading-order one-body contribution. 
The same cancellation has been observed in \cite{tree;bypark},
where $\Delta(1232)$ is accounted for only in the constants $c_2$ and
$c_3$ \cite{bernard-delta},
so that we
may consider that the role of the $\Delta(1232)$ is reasonably
reproduced at least at tree order. However, this may not be the
case in the one-loop order, since the energy scale of loops $p^0
\sim \sqrt{m_\pi m_N}$ is about the same size of the delta-nucleon
mass gap. Therefore, to be more satisfactory, the $\Delta(1232)$
should be included explicitly; we leave this extension for future work.

The most general form of the transition operator
effective at threshold reads \bea {\cal O}=(\tau_1^z+\tau_2^z)
\,(\vec{\sigma}_1-\vec{\sigma}_2)\cdot \left[ \vec{P}  A_{(1)}
 + \vec{k}  A_{(2)} P_{S=1} \right],
\label{eq;operators}
\eea
where
$\vec{P}=\vec{p}+\vec{p'}$ and
$\vec{k}=\vec{p}-\vec{p'}$, and
$\tau_j$ ($\vec{\sigma}_j$) is the isospin (spin) operator of
the $j$-th nucleon,
and $P_{S=1}= \frac{1}{4}(\vec{\sigma}_1 \cdot \vec{\sigma}_2+3)$
the spin-1 projection operator.
In deriving the above equation,
we have used
$ \vec\sigma_1\times\vec\sigma_2 = i (\vec\sigma_1-\vec\sigma_2) P^\sigma$,
where $P^\sigma=\frac12 (1 + \vec{\sigma}_1 \cdot \vec{\sigma}_2)$
is the exchange operator in spin space.

The $A_{(1)}$ receives contributions solely from the impulse diagram (a),
\be
A_{(1)}^{\rm IA} = \frac{g_A m_\pi}{8m_Nf_\pi}\,
 (2\pi)^3 \delta^{(3)}(\vec{k}).
\ee 
Note that all the higher order corrections at threshold are
already included in the above equation when the physical values of
$g_A$ and other parameters are used. On the other hand, the
$A_{(2)}$ receives contributions only from two-body graphs. The
contributions from the one-pion-exchange (OPE) diagrams including
the vertex loop corrections can be written as
\bea
A_{(2)}^{1\pi}
= \frac{g_A}{4f_\pi}\frac{1}{k^2-m_\pi^2} \Gamma_{\pi
N}(\vec{k},\vec{P}),
\eea
where $k^\mu\equiv
k_1^\mu=(k^0,\,\vec{k})$. Here and hereafter, we use the
FKA;
$k^0\equiv k_1^0=k_2^0=m_\pi/2$.
Up to one-loop accuracy, the $\Gamma_{\pi N}$ is given as
\bea
\lefteqn{\Gamma_{\pi N}(\vec{k},\vec{P}) =
\frac{2m_\pi^2}{f_\pi^2} \left[-2c_1+\frac{1}{2} \left(c_2+c_3
-\frac{g_A^2}{8m_N} \right)\right] }
\nnb \\ &&
+\frac{m_\pi}{m_Nf_\pi^2}\left[
\left(c_2+\frac{g_A^2}{32m_N}\right)\frac{m_\pi^2}{2}
+\left(c_2-\frac{g_A^2}{16m_N}\right) \vec{k}\cdot\vec{P}\right]
+ \frac{m_\pi^3}{32\sqrt{3}\pi f_\pi^4}
\nnb \\ &&
+ \frac{3g_A^2 m_\pi}{32\pi^2 f_\pi^4} \int^1_0dx \left[ x(1-x)
{\vec{k}}^2 +3 D(x,0,q-k) \right] n(x,0,q-k),
\label{eq;loop-pipinn}
\eea
with
\bea
n(x,\omega,k) &=&
\frac{m_\pi}{\sqrt{D(x,\omega,k)}} \left(\frac{\pi}{2}-{\rm
arctan}\frac{xk^0-\omega}{\sqrt{D(x,\omega,k)}} \right),
\label{eq;n0} \\
D(x,\omega,k) &=&
m_\pi^2-x(1-x)k^2-(xk^0-\omega)^2 \,.
\label{eq;D}
\eea
The four terms in Eq. (\ref{eq;loop-pipinn}) correspond to 
(b1), (b2), (b6), (b3) diagrams in Fig. \ref{fig;loops},
respectively.
Among the two-pion-exchange (TPE) graphs
drawn in Fig. 2($d1-d6$), only $(d1)$ and $(d2)$ contribute to
$A_{(2)}$:\footnote{One can easily distinguish the contributions
from $(d1)$ and $(d2)$ diagrams by noting that the contributions
of $(d1)$ and $(d2)$ are proportional to $g_A$ and $g_A^3$,
respectively.}
\bea
\lefteqn{
A_{(2)}^{2\pi} =
\frac{g_Am_\pi}{128\pi^2f_\pi^5}\int^1_0dx\left\{
-\frac{1-x}{2}n_-+\frac{x}{2}n_+
+ g_A^2
\left[  (9-10x)
{\rm ln}[1-x(1-x)\frac{k^2}{m_\pi^2}]
\right. \right.
}
\nnb \\ &&
+4(5x-6)\bar{n}_-+12\bar{n}_0-4(5x-3)\bar{n}_+
\left.\left.
-\frac{4x(1-x)\vec{k}^2}{m_\pi^2}
[(1-x)n_--n_0+x\,n_+] \right]\right\},
\eea
where $n_\pm =n(x,\pm m_\pi/2,-k)$, $n_0=n(x,0,-k)$, and similarly for the
functions with a bar, and 
\bea 
\bar{n}(x,\omega,k) &=&
\frac{D(x,\omega,k)}{m_\pi^2}n(x,\omega,k) . 
\label{eq;nb0} 
\eea
Finally we have the counter term contributions, drawn in Fig.
2$(c)$,
\bea A_{(1)}^{CT} =
\frac{g_Am_\pi}{128\pi^2f_\pi^5}(d_1^{(2)}-d_2^{(2)}
+d_3^{(2)}-d_4^{(2)}),
\ \ \ \ A_{(2)}^{CT} =
\frac{g_Am_\pi}{128\pi^2f_\pi^5}d_9^{(2)R}\,, 
\eea
with 
\bea
d^{(2)R}_9=d^{(2)}_9-4\,g_A^2 \left[
\frac{1}{\epsilon}
-\gamma 
+ {\rm ln}\left(\frac{4\pi \mu^2}{m_\pi^2}\right)
\right] ,
\eea
where 
$d^{(2)R}_9$ is the renormalized constant 
in ($4-2\epsilon$) dimension.
$\gamma=0.5772\cdots$ and $\mu$ is 
the renormalization scale of the dimensional regularization. 
In coordinate space, 
the counter term contributions are accompanied by the delta function. 
While the initial $pp$ state is in $P$-wave, 
these zero-ranged operators are
effective due to the derivative operators residing in Eq.
(\ref{eq;operators}). In principle, the $d^{(2)}$'s should be
determined from experiments, a task which is however not feasible
due to the lack of available data. We instead fix them from the
resonance-saturation assumption, taking into consideration the
omega and sigma meson exchanges \cite{tree;cohen}
\bea
A_{(1)}^{CT}
= -\frac{g_Am_\pi}{8f_\pi m_N^2}
\left(\frac{g_\sigma^2}{m_\sigma^2}
+\frac{g_\omega^2}{m_\omega^2}\right),
\ \ \ \
A_{(2)}^{CT} =
-\frac{g_Am_\pi}{4f_\pi m_N^2}
\frac{g_\omega^2}{m_\omega^2},
\eea
where $g_\sigma$ ($g_\omega$) and $m_\sigma$ ($m_\omega$) are the
scalar (vector) coupling and the mass of the sigma (omega) meson, 
respectively.
Using the values \cite{machleidt}
$g_\sigma=10.5$, $m_\sigma=508$ MeV, $g_\omega=10.1$, and
$m_\omega=783$ MeV, we have
$d_1^{(2)}-d_2^{(2)}+d_3^{(2)}-d_4^{(2)}=-7.70$ and
$d_9^{(2)R}=-4.78$.
Note that
we have imposed the zero-momentum subtraction scheme to the 
$A_{(2)}^{2\pi}$. 

The total cross section is given by
\bea
\sigma=\frac{1}{4}\frac{2\pi}{v_{\rm lab}}\int d\rho
 \left|T\right|^2 \ \ \ {\rm with} \ \ \
d\rho=\frac{2}{(2\pi)^4}m_p |{\vec q}|^2p_f d|{\vec q}|, 
\eea
where $v_{\rm lab}$ is the incident velocity in the lab frame, $T$
the transition amplitude, and $d\rho$ the phase factor; $m_p$ is
the proton mass, $p_f$ the magnitude of the relative
three-momentum of the final $pp$ state, and $\vec{q}$ the momentum
carried by the outgoing pion in the CM frame. The upper limit of
$|{\vec q}|$, $|{\vec q}|_{\rm max}$, is set by the initial total
energy.

In calculating the amplitude $T$, we use the wave functions
obtained by solving Schr\"{o}dinger equation with phenomenological
potentials. 
In this work we use the Reid soft core (RSC) and
Hamada-Johnston (HJ) potentials \cite{RSC,HJ}.\footnote{
In \cite{tree;cohen} it has been reported 
a large sensitivity
of the cross section to
the short-range part of the potential.
In our work we will study this feature 
by introducing a cutoff.
}
We write the initial
and final state wave functions as
\be
\Psi_i(r)=i \sqrt{2}\, \sqrt{4\pi}\,
 \frac{u_{1}(r)}{r} e^{i\delta_{1}}
|^{3}P_0\rangle,
\ \ \
\Psi_f(r)=\sqrt{4\pi}\, \frac{u_{0}(r)}{r} e^{i\delta_{0}}
|^{1}S_0\rangle,
\label{eq;s-wave} \ee
with the normalization condition
as $u_{L}(r)\stackrel{r\to \infty}{\longrightarrow}
\frac{1}{p} \sin(pr-\pi L/2 +\delta_{L})$.
The amplitude $T$ is given as
\bea
T &=& \frac{16\pi}{\sqrt{m_\pi}}
\int_0^\infty dr\, \left[
 \tilde A_{(1)} \left(
   u_0 u_1' - u_1 u_0' + \frac{2 u_0 u_1}{r} \right)
- \tilde A_{(2)} \frac{d}{dr} \left( u_0 u_1\right)
  \right],
\eea
where $u_L=u_L(r)$, $u_L'\equiv \frac{d}{dr} u_L(r)$,
and $\tilde A_{(j)}\equiv \tilde A_{(j)}(r)$ are the
transition operators in coordinate space,
\bea
\tilde A_{(j)}(r)=
 \int\frac{d^{3}\vec{k}}{(2\pi)^3}\,e^{i\vec{r}\cdot\vec{k}}
 A_{(j)}(\vec{k})\,.
\eea

\noindent {\bf 4. Numerical result and Discussion}

\begin{figure}
\begin{center}
\epsfig{file=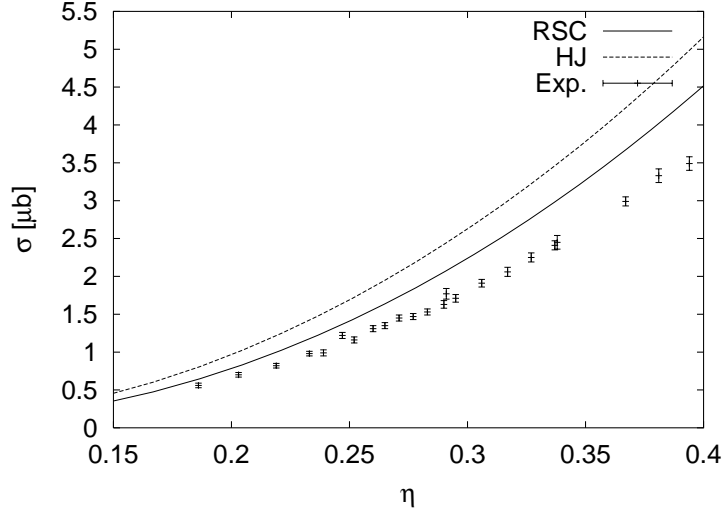,width=10cm} \caption{\it Total cross sections
up to the one-loop order in HBChPT. The wavefunctions are from RSC
(solid curve) and HJ (dashed curve) potentials. Experimental data
are also plotted.}
\end{center}
\label{fig;CS}
\end{figure}
In Fig. 3 we plot the total cross section vs. $\eta =
|\vec{q}|_{\rm max}/m_\pi$. Our theoretical predictions with loop
corrections come close to the experimental data. This
clearly points out the importance of the one-loop contribution.
This aspect has been noted before while mentioning
chiral-filter-unprotected cases.
Note that since the HJ potential has a hard-core whose radius is
about $0.5$ fm, the zero-ranged counter-term contributions are
identically zero. This corresponds to a ``hard-core
regularization" of the short-distance physics encoded in the
counter terms. This was referred to in \cite{chiralfilter2} as
``HCCS". For the RSC potential, the counter-term contributions
are not zero. To properly calculate the counter-term matrix
elements in a consistent regularization scheme, one has to resort
to what was referred to in \cite{chiralfilter2} as ``MHCCS" which
amounts to replacing the delta function and the two-body part by
\be
\delta(r) \rightarrow \delta(r-r_c),
\ \ \ \ \ \ \  
\tilde{A}_{(2)}^{1\pi,2\pi}(r)\to 
\theta(r-r_c) 
\tilde{A}_{(2)}^{1\pi,2\pi}(r) ,
\ee
The corresponding results with the RSC wavefunctions
are given in Fig. 4.
\begin{figure}
\begin{center}
\epsfig{file=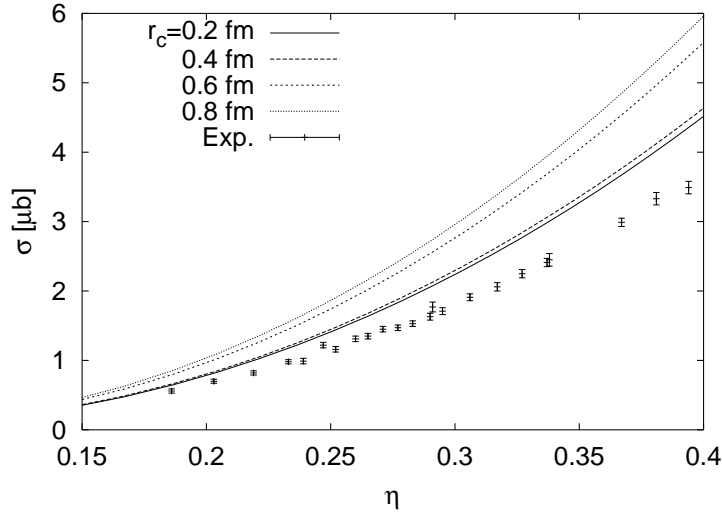,width=10cm}
\caption{\it 
Total cross section of our result 
for various cutoff $r_c$ with the wave-function from RSC potential.
The contributions from the
zero-ranged operators are also included with
replacing $\delta(r)$ by $\delta(r-r_c)$.
}
\label{fig;CS-CT-reid}
\end{center}
\end{figure}
As can be seen in Fig. 4, the results are mildly sensitive to
$r_c$. 
This mild sensitivity -- which may be taken as a justification
of the short-distance regularization procedure used here -- was
also observed in \cite{chiralfilter2,PKMRhep} in different
contexts.

A serious question here is the convergence of the perturbation
series. It is difficult to see whether or not the series actually
converges without computing higher order terms. But going to
higher order is not feasible at the moment. To gain a rough idea
from the present calculation, we may consider the break-down of
the transition amplitude into individual contributions. For
convenience of comparison, we measure correction terms relative to
the IA amplitude calculated with the RSC wavefunctions at
$|\vec{q}|_{\rm max}= 0.1\ m_\pi$. Then the leading MEC (coming
from OPE) is $-0.8$ in units of the IA term, leading to a
substantial cancellation. The one-loop order MEC contribution is
$(-1.0) + (-1.2)= (-2.2)$, where the first (second) term 
represents the individual contribution from OPE (TPE) diagrams.
Summing up the whole (including IA) contribution, we
get $-2.0$, that is, twice the IA contribution with the
opposite sign. 
This shows that there is no convergence to the
order considered. 
This feature is shared by other
chiral-filter-unprotected cases; 
in both the polarized {\it np}-capture 
and the ``hep" process, a similar pattern is seen.
There is however one important difference. 
In the latter cases,
the coefficients of the counter terms can be reduced to one
effective constant 
and, 
when this constant is fixed by experiments,
can be more or less accounted for higher order terms
that are not computed 
(such as the magnetic moment of the deuteron in the case of
polarized {\it np}-capture and the triton beta decay in the case
of ``hep"). This amounts to terminating the series by fiat at the
order considered by the counter terms. One can think of this as
choosing -- as one does in gauge theory calculations -- an optimal
regularization scheme so as to account for higher-order terms that
are not computed explicitly.
This freedom of choosing the optimal
regularization scheme is not available in the $\pi^0$ production
case since the counter term is entirely fixed 
by means of the resonance saturation 
and hence cannot ``mock up" higher-order terms which
may not be negligible.

Next, we discuss the effect of ISI and FSI in studying the
convergence of the $1/m_N$ correction of the $c_2$ term. 
In \cite{bernard} the $1/m_N$ contribution was reported to become 
roughly twice as large as the leading $c_2$ term;
this result was obtained by employing a
simple ``factorization ansatz" 
which dictates that the nucleon legs of the
production operator are imposed to be on mass-shell. 
This is
confirmed in \cite{loop;DKMS}: The $1/m_N$ correction, (b2)
diagram in Fig. \ref{fig;loops}, is about 260 \% of the leading
MEC contribution (b1) when the ISI and FSI are ignored. 
We would obtain
the same result if we exclude the ISI and FSI, 
that is, imposing the
on-shell condition $\vec{k}\cdot\vec{P}= m_\pi m_N+m_\pi^2/4$ 
in Eq. (\ref{eq;loop-pipinn}). 
However, after taking account of the
ISI and FSI we find the matrix element of (b2)
is only about 20 \% and 40 \% of the
(b1) evaluated with RSC and HJ potential, respectively.
Consequently, the $1/m_N$ correction does not show any anomalous
behavior in our calculation. The similar pattern of softening the
$1/m_N$ correction of the Weinberg-Tomozawa term due to the ISI
and FSI was observed in \cite{rocha} as well.

We want to remark on Sato {\it et al.}'s observation
\cite{momentum}: In a momentum space calculation with the
half-off-shell wavefunctions, they found significant contributions
from the momentum region even larger than the chiral scale
$\Lambda_\chi\sim 1$ GeV for some Feynman diagrams. Regarding this
apparent disagreement between their cutoff-dependence and our
$r_c$-independence, we should note that the observed
$r_c$-independence in our theory is in the total amplitude, not in
individual terms. In fact if we evaluate the contribution for each
diagram, a considerable $r_c$-dependence (though not as radical as
theirs) is observed in our calculation also.

In summary, we confirm quantitatively that the two pion exchange
one-loop contribution is quite important. 
We also find that there
is no strong dependence on the short range interaction. 
This result suggests
that massive degrees of freedom (not included explicitly
in our study) may have been properly incorporated into the
diagrams and counter terms investigated in our study. 
The apparently poor HBChPT convergence however may indicate that
significant contributions from higher-order operators may be left
unaccounted for in our scheme. Furthermore, the validity of the
FKA should be examined.  
Clearly more work is required 
including comparison with the other counting rules\cite{tree;cohen,
loop;gedalin}, studies of next order terms,
and the verification of the approximations involved.

\section*{Acknowledgement}

We thank M. Rho, K. Kubodera, F. Myhrer and the nuclear theory
group of University of South Carolina for useful discussions. SA
also thanks B.-Y. Park for discussions. This research was
supported in part by the KOSEF Grant 985-0200-001-2 and KRF
1999-015-DI0023, and by the National Science Foundation, Grant No.
PHY-9900756 and No. INT-9730847.

\end{document}